\documentclass[12pt]{iopart}
\newcommand{\gguide}{{\it Preparing graphics for IOP journals}}
\usepackage{iopams}
\usepackage{graphicx}
\usepackage{hyperref}

\begin{document}
\title[Author guidelines for IOP journals in  \LaTeXe]{Spectroscopy and decay properties of $\Sigma_{b}, \Lambda_{b}$ baryons in quark-diquark model}
\author{Ajay Majethiya$^{*}$ ,\ Kaushal Thakkar and P C Vinodkumar}
\address{Kalol Institute of Technology and Research centre, Kalol-382 721, India,\\Department of Physics, Sardar Patel University, Vallabh Vidyanagar 388 120,\\ Gujarat, India \\
} \ead{ajay.phy@gmail.com}
\begin{abstract}
Properties of single beauty baryons ($\Sigma_{b},\Lambda_{b}$) are studied based on the quark-diquark structure.
The confinement potential is assumed as two body colour coulomb plus power potential with exponent $\nu$.
We find a strong correlation between the choice of the heavy quark mass parameter ($m_{b}$),
strong coupling constant ($\alpha_{s}$) and the potential exponent ($\nu$) for getting the experimental mass spilt  of
 $m_{\Sigma^{*}_{b}}- m_{\Sigma_{b}}=21.2\pm2.0 $ MeV. The resultant spectroscopic parameters are used for computing
 magnetic moments, the electromagnetic radiative decay, strong hadronic decay and semileptonic decay widths of $\Sigma_{b}, \Lambda_{b}$ systems. Our predictions on the radiative decay width correspond to $\Sigma^{*0}_{b}\rightarrow\Lambda_{b}\gamma$  are in agreement with the QCD sum rule prediction. The present results on the semileptonic decay widths of $\Lambda_{b}\rightarrow \Lambda_{c}l\nu_{l}$ (2.50-4.73) $ *10^{10} s^{-1}$
are in agreement with the experimental value of $3.59^{+1.234}_{-0.936} *10^{10} s^{-1}$ reported by (PDG 2010).

\end{abstract}
\maketitle
\def\be{\begin{equation}}      \def\ol{\overline}
  \def\ee{\end{equation}}    \def\beq{\begin{eqnarray}}
  \def\dis{\displaystyle}    \def\eeq{\end{eqnarray}}
  \def\btd{\bigtriangledown}     \def\m{\multicolumn}
  \def\dfac{\dis\frac}       \def\ra{\rightarrow}

\section{Introduction}
Study of the heavy flavour hadrons has become a
subject of renewed interest due to recent observations reported by the experimental groups at
Belle, BABAR, DELPHI, CLEO, CDF etc;
\cite{Mizuk2005,Aubert2007,Aaltonen07,Feindt2007,Artuso2001}. Most of the new states are observed within the heavy flavour sector with one or more heavy flavour quark composition. Many phenomenological models
have also predicted the heavy flavour baryon masses.
Capstick and Isgur studied the heavy baryon system in a
relativized quark potential model \cite{Capstick1986}. Roncaglia et al.
predicted the masses of baryons containing one or two
heavy quarks using the Feynman-Hellmann theorem and
semiempirical mass formula \cite{Roncaglia1995}. Mathur et al. predicted the masses of charmed
and bottom baryons from lattice QCD \cite{Mathur2002}. Ebert et al.
calculated the masses of heavy baryons in the lightdiquark
approximation \cite{Ebert:2005}. Using the relativistic
Faddeev approach, Gerasyuta and Ivanov calculated the
masses of the S-wave charmed baryons \cite{GI1999}. Later Gerasyuta and Matskevich studied the charmed
baryon multiplets using the same approach \cite{GM2000}.
It
is expected that more states of heavy flavour baryons will be detected
in coming years. Though there are
consensus among the theoretical predictions on the ground state
masses \cite {Giannini2001,Ebert2005}, there seemed to have little
agreement among the model predictions of the properties like
the mass difference among the different spin-parity of baryonic states, the form factors
\cite{Giannini2001}, magnetic moments \cite{Bhavin2008} etc;. Stimulated by recent experimental progress, the study of the heavy flavour spectroscopy
is becoming extremely rich and interesting.\\
Due to the rich mass spectrum and the relatively narrow widths, the
heavy baryon system provides an excellent ground for testing the ideas and predictions of heavy
quark symmetry and light flavour symmetry. The pseudoscalar mesons involved in the strong
decays of charmed baryons are soft. By studying the strong decay modes, one expects to extract information about their structures and the low energy dynamics of heavy baryons $\emph{vis a vis}$ interaction with pion and other pseudoscalar mesons. \\
Semileptonic decay of hadrons are of interest for two basic reasons; they are the primary source of information for the extraction of the Cabbibo-Kobayashi-Maskawa (CKM) matrix elements of the Standard Model from
experiment, and the study of the semileptonic decays of baryons provides information about the nature of the interquark interactions \cite{muslema2006} while it undergoes transformations and it supplements similar
informations gathered from the heavy flavour meson decays. These processes also act as good probes
for the factorization hypothesis which has been extensively
explored for dealing with hadronic transitions \cite{Ali1998, Chen1999}.
Simultaneously in recent times, many semileptonic and nonleptonic decays of
$\Lambda_{b}$ are experimentally recorded \cite{K. Nakamura2010, Abdallah2004}. Moreover the
LHCb is expected to accumulate a large data sample of
b hadrons with enhanced luminosity factor that offers unique opportunities for studying many new hadronic states and their decay properties.\\ The first bottom baryon $\Lambda _b $ (udb), within mass around 5640 MeV, was reported by UA1
Collaboration at CERN in late 1990s \cite{Albajar}. Later the
$\Lambda _b $ was confirmed by other experimental groups such as DELPHI
Collaboration \cite{Abreu}, ALEPH Collaboration \cite{Buskulic},
and CDF Collaboration \cite{Abe} within the mass range 5614 to 5668 MeV. Recently, the mass of $\Lambda _b $ was
further measured to be 5619.7 MeV by the CDF Collaboration at
Fermilab \cite{Acosta}. Very recently five new bottom baryons,
$\Sigma_b^{(*)}$ and $\Xi_b^-$ were reported by the CDF Collaboration at
Fermilab \cite{CDF,Xi0} in proton-antiproton collisions at $\sqrt s
$=1.96 TeV.\\
The present paper is thus aimed to study both the static and dynamic properties of $\Sigma_{b}, \Lambda_{b}
$ baryons in the quark-diquark model for baryons by considering light quarks
for the construction of diquark
states.\\

\section{Theoretical framework}
Following Gell-Mann's suggestion of the possibility of quark-diquark stucture for baryons \cite{M.
Gell-Mann1964}, various authors have introduced effective degrees of freedom of diquarks in order to
describe baryons as composed of a constituent diquark and quark \cite{M. Ida1966, D.
B. Lichtenberg1967,M. Anselmino1993,F.Wilczek2005}. The presence of a coherent diquark
structure within baryons helps us to treat the problem of three-body to that of two two-body interactions. In
the present study of heavy flavour baryons containing one beauty qaurk, it is appropriate to consider the two light quarks as the diquark states. \\
 \begin{figure*}
\begin{center}
\includegraphics[height=3.0in,width=2.9in]{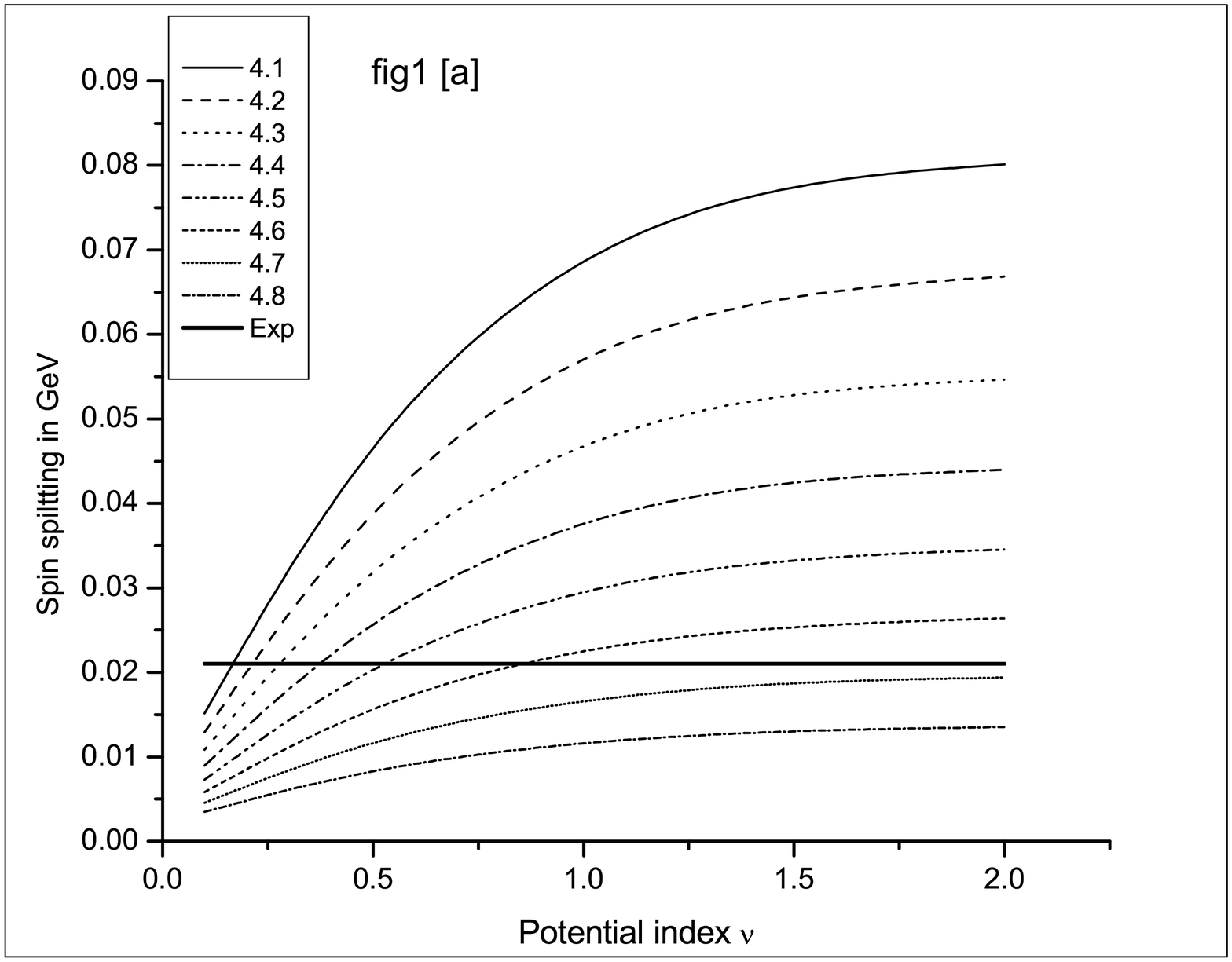}
\includegraphics[height=3.0in,width=2.9in]{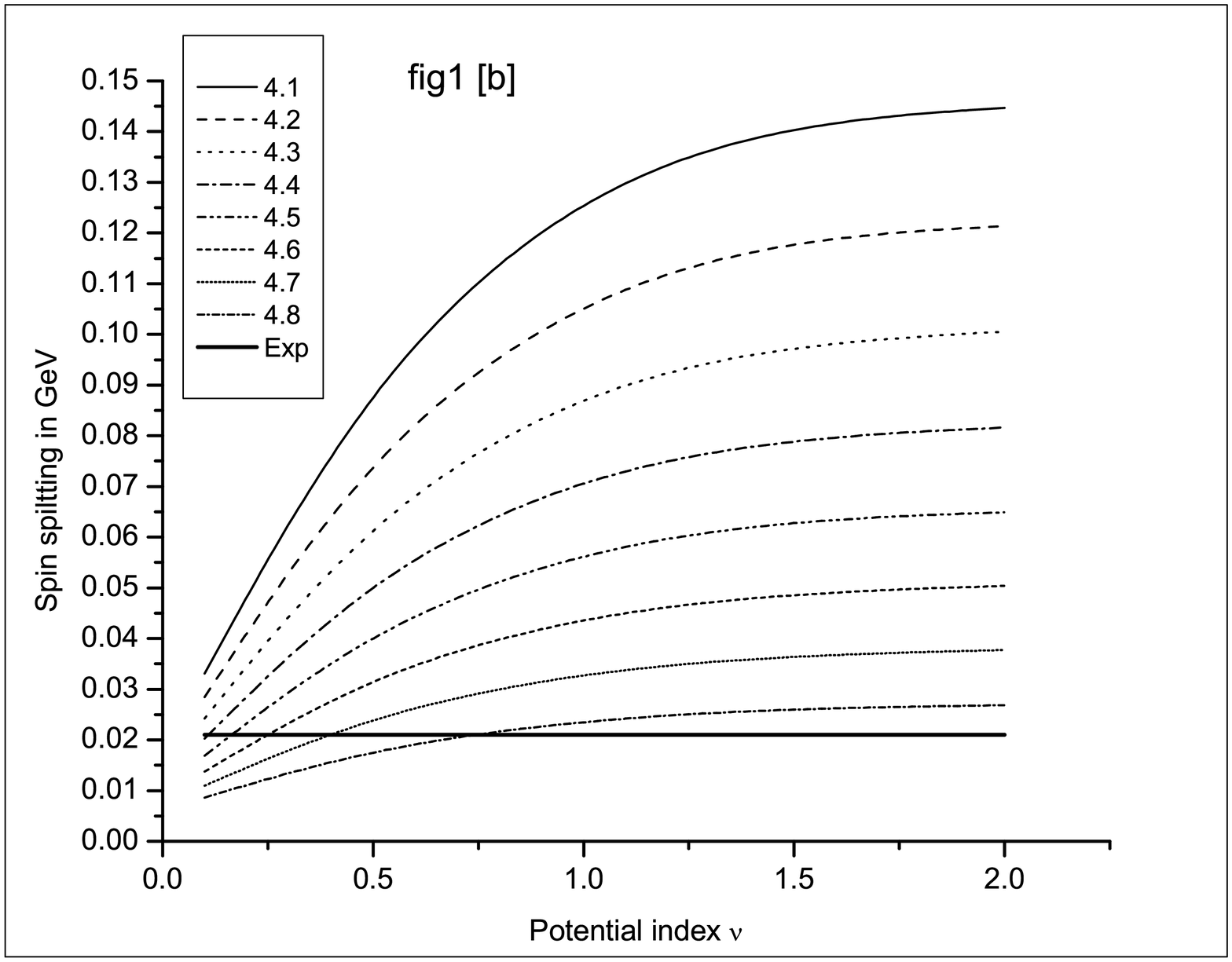}\vspace{0.1in} \vspace{0.1in}
\includegraphics[height=3.0in,width=2.9in]{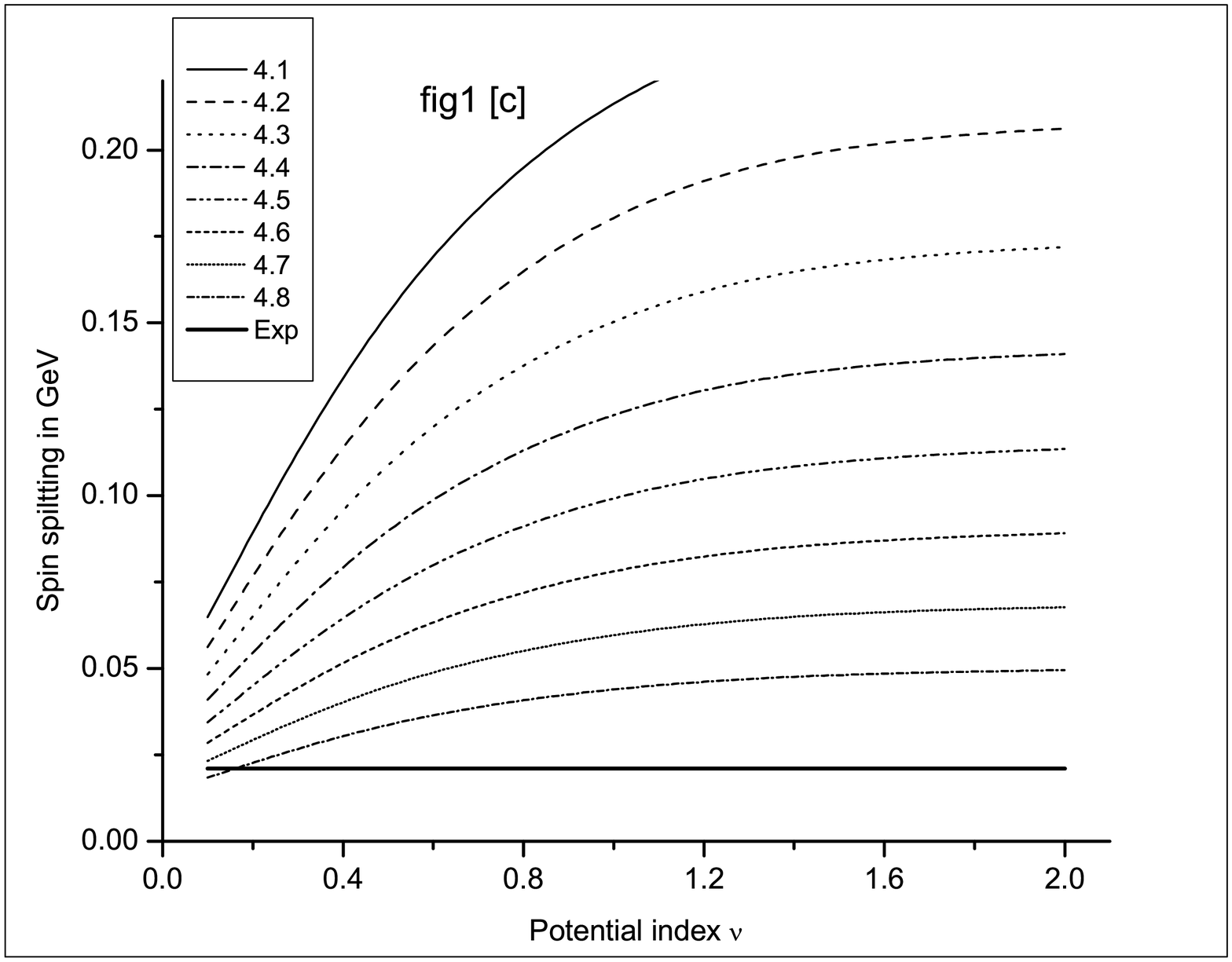}
\includegraphics[height=3.0in,width=2.9in]{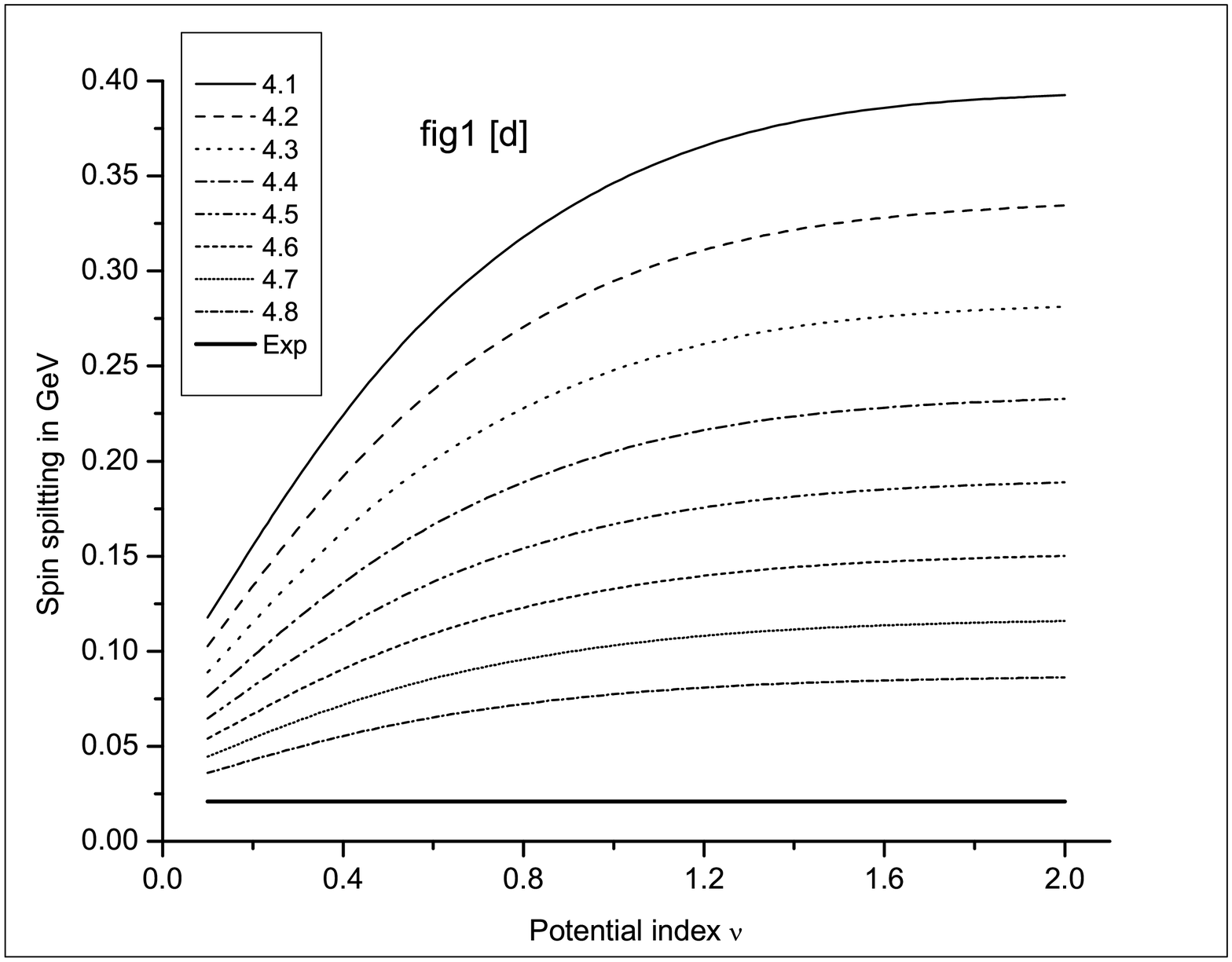}
\vspace{0.1in} \vspace{0.1in}

\caption{\textbf{Variation of spin spiltting
with potential index $\nu$ with different range of running strong coupling constant, $\alpha_{s}$ for single beauty baryons, $\alpha_{s}=0.15$ [a], $\alpha_{s}=0.20$ [b], $\alpha_{s}=0.25$ [c], $\alpha_{s}=0.30$[d].}}\label{fig:massvspot}
\end{center}
\end{figure*}
\\ In the quark-diquark model, the Hamiltonian of the baryon is expressed in terms of a
diquark Hamiltonian ($H_{jk}$) plus qurk-diquark Hamiltonian ($H_{i,jk}$) as \cite{W. S. Carvalho1994}
\begin{equation}\label{eq:1}
 H=H_{jk}+H_{i,jk}
\end{equation}
Here, the internal motion of the diquark($jk$) is described by
\begin{equation}\label{eq:2}
H_{d}=H_{jk}=\frac{{p}^2}{2m_{jk}}+V_{jk}(r_{jk})
\end{equation}
 and the Hamiltonian
of the relative motion of the diquark($d$)- quark($i$) system is described by
\begin{equation}\label{eq:3}
H_{i,d}=H_{i,jk}=\frac{{q}^2}{2m_{i,jk}}+V_{i,jk}(r_{id})
\end{equation}
where, $p$ is the relative momentum of the quarks within the diquark and q is the relative momentum of the quark-diquark system. The reduced mass of the two body systems appeared in Eq.\ref{eq:2} and Eq.\ref{eq:3} respectively are defined as
\begin{equation} m_{jk}=\frac{m_{j} m_{k}}{m_{j}+m_{k}},\\ m_{i,jk}=\frac{m_{i}(m_{j}+ m_{k})}{m_{i}+m_{j}+m_{k}}
\end{equation}\label{eq:4}
For the present study, we have assumed color coloumb plus power potential for the interquark potential of Eq.\ref{eq:2} as well as for the the quark-diquark interaction of Eq.\ref{eq:3}. \\Accordingly, the diquark potential is written as,
\begin{equation}\label{eq:05}
V_{jk}=-\frac{2}{3}\alpha_s\frac{1}{r_{jk}}+ b{\,\,}  r^{\nu}_{jk}
\end{equation}
and the quark-diquark potential as
\begin{equation}
V_{i,jk}=-\frac{4}{3}\alpha_s\frac{1}{r_{id}}+ b{\,\,}  r^{\nu}_{id}
\end{equation}
where, $r_{id}$ is the quark-diquark separation distance, $\nu $ is the exponent
corresponding to the confining part of the potential and $b$ is the strength
of the potential, which is assumed to be same for the di-quark interaction as well as
between the quark-diquark interaction. The Schrodinger equation corresponds to the Hamiltonian given in Eq.\ref{eq:1}. is numerically solved using the Runge-Kutta method in a mathematica note book. The degeneracy of the states are removed by introducing the spin dependent
interaction potential given by \cite{S. S. Gershtein2000} \begin{eqnarray}
V^{(d)}_{SD}(r_{jk}) &=& \frac{2}{3}\alpha_s
\frac{1}{3 m_{j} m_{k}}{S_{j} \ \cdot S_{k}} [4\pi \delta(r_{jk})]\end{eqnarray}\\
among the diquark states, and
\begin{eqnarray}\label{eq:2.10}
V^{(i-d)}_{SD}(r)&=& \frac{4}{3}\alpha_s \frac{1}{3 m_{i} 2 m_{jk}}({S_{d}+ L_{d}) \ \cdot S_{q}} [4\pi
\delta(r_{id})]
\end{eqnarray}
among the quark-diquark($id$) system. \\
 The potential parameters of the model are fixed to yield the spin average mass
 as well as the hyperfine spiltting of the ground state of $\Sigma^{*}_{b}(5829)-\Sigma_{b}(5808)$ baryon.
 The mass spiltting has been studied for different choices of the quark mass parameter,
 $m_{b}$ for each case of the potential exponent ($\nu$) with
 the different choice of the running strong coupling constant $\alpha_{s}$. The trendlines shown in fig 1 (a-d) are the predicted masss spilts against
 the potential exponent with different choices of $\alpha_{s}$ and $m_{b}$. The solid horizontal line drawn in these plots correspond
 to the experimental mass spilt of 21 MeV. It is found that the choices of heavy quark mass parameter, $m_{b}$ and
 $\alpha_{s}$ play a decisive role in the mass splitting of the ground state.
The plots in fig 1 tend to saturate beyond the potential exponent $\nu > 1$. From the plots in Fig.1 [a-d], it is quite clear that there is a very restricted parameter space in the choices of $\alpha_{s}$ , $m_{b}$ and the potential exponent $\nu$, that provide the experimental mass spilt of 21 MeV.
These model parameters thus extracted here for the choices of $\alpha_{s}=0.15$ (Set A) and $\alpha_{s}=0.20$
(Set B) are listed in Table 1. With these sets of input values, the predicted ground state mass of the $\Lambda_{b}$ baryons for the different choices of $\nu$ are listed in Table \ref{tab:1}. The two parameter sets (A and B) deduced from the spectroscopy are now be employed to compute the magnetic moments, radiative decay, strong decay
and semileptonic decay widths of  $\Sigma_{b}$ and  $\Lambda_{b}$ baryons with no additional parameters.\\
\begin{table}\caption{Masses (in GeV) of $\Lambda_{b}$ with other model parameters.} \vspace{0.01in}
\begin{center}\label{tab:1}
\begin{tabular}{lcccc}
\hline
$\alpha_{s}$&$\nu$ &$b$&$m_{b}$&$M_{\Lambda_{b}}$\\
\hline
	        &	0.4	&	0.1601	&	4.450	&	5.660	\\
        	&	0.6	&	0.0940	&	4.550	&	5.657	\\
	0.15    &	0.8	&	0.0595	&	4.600	&	5.655	\\
  (Set A)  	&	1.0	&	0.0389	&	4.630	&	5.654	\\
\hline										
	    	&	0.2	&	0.1968	&	4.570	&	5.656	\\
	 	    &	0.4	&	0.0976	&	4.720	&	5.651	\\
	0.20  	&	0.6	&	0.0563	&	4.780	&	5.647	\\
   (Set B)   &	0.8	&	0.0341	&	4.815	&	5.647	\\
	      	&	1.0	&	0.0211	&	4.840	&	5.647	\\
\hline
	PDG\cite{K. Nakamura2010}	&	-	&	-	&&	5.620$\pm$0.0016	\\
\hline
\end{tabular}
\end{center}
\end{table}
\begin{table}\caption{\label{tab:02} Magnetic moments  of the $\Sigma_{b}$ and
$\Lambda_{b}$  ($J^{P}=\frac{1} {2}^{+}$ Baryons) in terms of nuclear magneton $\mu_{N}$.}
\begin{center}
\begin{tabular}{lr|rr|rr|rr|rrrrr}
\hline
Parameters&&$\Sigma^{+}_{b}$&&$\Sigma^{0}_{b}$&&$\Sigma^{-}_{b}$&
&$\Lambda^{0}_{b}$&\\
 Set&$\nu$&cqm&ecqm&cqm&ecqm&cqm&ecqm&cqm&ecqm\\
\hline

  &0.4 &2.505&2.210&0.643&0.568&-1.217&-1.074&-0.070&-0.063\\
  &0.6&2.505&2.253&0.643 &0.578&-1.218&-1.095&-0.069&-0.063\\
 A  &0.8&2.504&2.274&0.643&0.584&-1.218&-1.106 &-0.068&-0.063 \\
  &1.0&2.504&2.287&0.643&0.587&-1.128&-1.112&-0.067&-0.063\\
\hline
 &0.2&2.505&2.262&0.643&0.580&-1.218&-1.100& -0.068&-0.063 \\
 &0.4&2.504&2.326&0.642&0.596&-1.219&-1.132& -0.066& -0.063\\
  B&0.6&2.504&2.351&0.642&0.603&-1.219&-1.145&-0.065&-0.063 \\
 &0.8&2.503&2.366&0.642&0.607&-1.219&-1.152&-0.064 &-0.063\\
&1.0&2.503&2.377&0.642&0.609&-1.219&-1.158&-0.064&-0.063 \\
\hline
\cite{Bhavin2008}      &&2.226&-&0.591&-&-1.045&-&-0.064&- \\
\cite{B.Silvestre-brac1996}&      &2.669&-&0.682&-&-1.305&-&-0.060&- \\
RQM\cite{Faessler 2006}     & &2.070&-&0.530&-&-1.010&-&-0.069&- \\
NRQM\cite{Faessler 2006}     & &2.010&-&0.520&-&-0.970&-&-0.060&- \\
\hline

\end{tabular}
\end{center}
\end{table}

\begin{table}\caption{\label{tab:03} Magnetic moments  of the $\Sigma_{b}$ and
$\Lambda_{b}$ ($J^{P}=\frac{3} {2}^{+}$ Baryons) in terms of nuclear magneton $\mu_{N}$.}
\begin{center}
\begin{tabular}{lr|rr|rr|rr}

\hline
 Parameter &&$\Sigma^{+}_{b}$&&$\Sigma^{0}_{b}$&&$\Sigma^{-}_{b}$&\\
\hline
Set&$\nu$&cqm&ecqm&cqm&ecqm&cqm&ecqm\\
\hline
   &0.4&3.653&3.211&0.862&0.756&-1.930&-1.698\\
  &0.6&3.654&3.275&0.862 &0.772&-1.930&-1.730\\
 A  &0.8&3.655&3.307&0.862&0.780&-1.930&-1.746 \\
  &1.0&3.655&3.326&0.862&0.785&-1.929&-1.755\\
\hline
 &0.2&3.654&3.288&0.862&0.775&-1.930&-1.736 \\
&0.4&3.657&3.384&0.864&0.800&-1.927&-1.784 \\
B &0.6&3.657&3.423&0.865&0.809&-1.927&-1.803 \\
 &0.8&3.658&3.445&0.865&0.815&-1.927&-1.814 \\
 &1.0&3.658&3.461&0.866&0.819&-1.926&-1.822 \\
 \hline
\cite{Bhavin2008}     & &3.239&-&0.791&-&-1.655& \\
\cite{T. M. Aliv2009}  &    &2.50$\pm0.50$&-&0.50$\pm0.15$&-&-1.50$\pm0.36$& \\
\hline

\end{tabular}
\end{center}
\end{table}

\section{Magnetic Moments of the $\Sigma_{b}$ and
$\Lambda_{b}$ Baryons}
The magnetic moment of a baryon is computed in terms of its constituent quarks as
\cite{Bhavin2008}
\begin{equation}
\mu_B=\sum\limits_{i}\left<\phi_{sf}\mid\mu_{i}
\overrightarrow{\sigma}_{i}\mid\phi_{sf}\right>
\end{equation}
where, $\left|\phi_{sf}\right>$ represents the spin-flavour
wave function of the quark composition constituting the baryonic state \cite{Bhavin2008},
$\mu_{i}$ is expressed as
\begin{equation}
\mu_{i}=\frac{e_{i}}{2m_{i}}
\end{equation}
in terms of the charge, $e_{i}$ and the mass of the bound quarks, $m_{i}$, while $\sigma_{i}$ represents the spin of the respective constituent quark
corresponding  to the spin flavour wavefunction of the baryonic state. Apart from the model mass parameter($m_{i}$)
of the quarks, it would be appropriate to define an effective mass ($m^{eff}_{i}$) of the bound quarks confined within the hadron as \cite{Bhavin2008}

\begin{equation}
m^{eff}_{i}=m_i\left( 1+\frac{<H>}{\sum\limits_{i}m_{i}}\right)
 \end{equation}
such that the baryon mass is given by
\begin{equation}
M_B=\sum\limits_{i}m^{eff}_{i}
\end{equation}
The magnetic moments are now computed using the model constituent quark mass (cqm) and also by considering
effective mass of the bound quarks (ecqm). Present results of the magnetic moments in terms of the nuclear magnetons ($\mu_{N}$) of $J=\frac{1}{2}$ ($\Sigma^{\pm,0}_{b}$, $\Lambda^{0}_{b}$) and $J=\frac{3}{2}$ ($\Sigma^{*}_{b}$) states of the beauty baryons are presented in
Table 2 and 3 respectively.\\
\begin{table*}
\begin{center}
\caption{\label{tab:04}Radiative decay widths ($\Gamma_{\gamma}$) of beauty baryons in terms of keV (* indicates $J^P=\frac{3}{2}^+$
state.)}
\begin{tabular}{lrrrrrrrr}
\hline Decay&$\nu$ &$\mu$ (in $\mu_{N}$)&{\,\,\,\,\,\,\,\,\,\,\,\,\,\,\,\,}$\Gamma_{\gamma}$&&Others\\
\hline &
&&(cqm)&(ecqm)&&\\
\hline
$\Sigma^{*0}_{b}\rightarrow\Lambda^{0}_{b}\gamma$&0.4&2.03&98.64&78.19&114.00\cite{Aliev2008}\\
&0.6&2.07&104.08&85.79&344.00\cite{Zhu1999}\\
Set A&0.8&2.09&107.82&90.58&251.00\cite{Stawfiq2001}\\
&1.0&2.11&109.72&93.97&\\
\cline{2-5}
&0.2&2.08&105.94&88.33&\\
&0.4&2.14&115.57&101.99&\\
Set B&0.6&2.16&123.67&111.71&\\
&0.8&2.18&123.67&113.16&\\
&1.0&2.19&123.67&114.20&\\
\hline
$\Sigma^{0}_{b}\rightarrow
\Lambda^{0}_{b}\gamma$&0.4&-1.440&65.74&52.44&\\
&0.6&-1.469&69.90&58.03&\\
Set A&0.8&-1.483&72.77&61.61&\\
&1.0&-1.492&74.24&63.58&\\
\cline{2-5}
&0.2&-1.474&71.33&59.69&\\
&0.4&-1.517&78.75&69.80&\\
Set B&0.6&-1.535&85.05&77.12&\\
&0.8&-1.545&85.05&78.10&\\
&1.0&-1.552&85.05&78.81&\\
\hline
\end{tabular}
\end{center}
\end{table*}
\section{Radiative decay of beauty Baryons}
\begin{table*}[t]
\begin{center}
\caption{\label{tab:05}Radiative decay widths ($\Gamma_{\gamma}$) of singly
charmed baryons in terms of keV (* indicates $J^P=\frac{3}{2}^+$
state.)}
\begin{tabular}{lrrrrrr}
\hline Decay&$\nu$&$\mu$ (in $\mu_{N}$) &&$\Gamma_{\gamma}$ in MeV&Others\\
\hline &
&&(cqm)&(ecqm)&\\
\hline

$\Sigma^{*-}_{b}\rightarrow \Sigma^{-}_{b}\gamma$&0.4&-0.714&0.0259&0.0195&0.11\cite{Aliev2008}\\
&0.6&-0.729&0.0258&0.0204&\\
Set A&0.8&-0.737&0.0256&0.0208&\\
&1.0&-0.742&0.0255&0.0211&\\
\cline{2-5}
&0.2&-0.732&0.0254&0.0206&\\
&0.4&-0.755&0.0253&0.0219&\\
Set B&0.6&-0.764&0.0252&0.0224&\\
&0.8&-0.770&0.0252&0.0227&\\
&1.0&-0.774&0.0251&0.0230&\\
\hline
$\Sigma^{*+}_{b}\rightarrow
\Sigma^{+}_{b}\gamma$&0.4&1.604&0.127&0.098&0.46\cite{Aliev2008}\\
&0.6&1.634&0.127&0.102&1.26\cite{Zhu1999} \\
Set A&0.8&1.649&0.127&0.104&\\
&1.0&1.658&0.127&0.105&\\
\cline{2-5}
&0.2&1.640&0.127&0.103&\\
&0.4&1.685&0.126&0.109&\\
Set B&0.6&1.703&0.126&0.111&\\
&0.8&1.713&0.126&0.112&\\
&1.0&1.721&0.126&0.113&\\
\hline
 $\Sigma^{*0}_{b}\rightarrow
\Sigma^{0}_{b}\gamma$&0.4&0.444&0.010&0.0076&0.03\cite{Aliev2008}\\
&0.6&0.452&0.0098&0.0078&0.08\cite{Zhu1999}\\
Set A&0.8&0.455&0.0096&0.0079&0.15\cite{Stawfiq2001}\\
&1.0&0.458&0.0095&0.0080&\\
\cline{2-5}
&0.2&0.453&0.0096&0.0079&\\
&0.4&0.464&0.0095&0.0082&\\
Set B&0.6&0.467&0.0094&0.0083&\\
&0.8&0.471&0.0093&0.0085&\\
&1.0&0.473&0.0092&0.0086&\\
\hline
\end{tabular}
\\
 \vspace{0.1in} \cite{Stawfiq2001} $\rightarrow$ HQS, \cite{Aliev2008,Zhu1999}$\rightarrow$ QCD SUM RULES,

\end{center}
\end{table*}
The  electromagnetic radiative decay of beauty baryons have been computed by various approaches such as
in the leading order of Heavy Quark Effective Theory (HQET)\cite{Zhu1999}, heavy quark and chiral symmetries \cite{Stawfiq2001},  light cone QCD sum rules \cite{Aliev2008} etc;. However, there exist wide disparities among these predicted values. So it is important to predict the radiative decay width along with other properties of the beauty baryons within a single scheme. \\
In terms of the radiative transition magnetic moment,
the electromagnetic radiative decay width can be computed as \cite{JDey1994}
\begin{equation}
\Gamma_{\gamma}=\frac{k^3}{4\pi}\frac{2}{2J+1}\frac{e^2}{m_{p}^2}
\mu^{2}_{B\rightarrow B^{'} \gamma}
\end{equation}
here, $m_{p}$  is the proton mass, $\mu_{B\rightarrow B^{'} \gamma}$ is the radiative transition
magnetic moments (in nuclear magnetons), which are expressed in
terms of the magnetic moments of the constituting quarks ($\mu_{i}$)
of the initial (B) and final ($B^{'}$) states of the baryon \cite{JDey1994} and k is the photon energy.
The radiative transition magnetic moment corresponds to  $B \rightarrow
B'\gamma$ can be computed in terms of the spin-flavour wave functions
of B' and B states as
\begin{equation}
\mu_{B \rightarrow B'\gamma} = \sum\limits_{i}\left<\phi_{sf}^
{B'}\mid\mu_{i} \vec{\sigma}_{i}\mid\phi_{sf}^ {B}\right>
\end{equation}
here, $\mu_{i}$ is as given by Eq.(10). However,
 the effective mass of the bound quarks of the $B' - B$ system is defined in terms of the respective mass of the bound quarks constituting the $B'$ and $B$ states as
\begin{equation}
m_{i}^{eff} = \sqrt{m^{eff}_{i(B)} m^{eff}_{i(B')}}
\end{equation}
Using the spin flavour wave function of the $B'$ and $B$
states, transition magnetic moment is
computed the model quark mass parameters (cqm) as well as by considering the quark
confinment effect through the effective mass of the bound quarks (ecqm).

The predicted transition magnetic moments and radiative decay widths corrosponds,
$(\Sigma^{*0}_{b},\Sigma^{0}_{b})\rightarrow\Lambda_{b}\gamma$ and $\Sigma^{*}_{b}\rightarrow \Sigma_{b}\gamma$ transitions are listed in Table \ref{tab:04} and \ref{tab:05} along with other model predictions.
\section{Hadronic strong decay width of beauty Baryons}
By studying the strong decay modes, one expects to extract information about their structures and the low energy dynamics of heavy baryons $\emph{vis a vis}$ interaction with pion and other pseudoscalar mesons. Many theoretical models have studied the strong decay of heavy baryons in different formalism. Xin-Hen Guo et. al,
 has predicted the strong decay width in the Bethe-Salpeter formalism \cite{Xin2008}. Hai-Yang Cheng et.al,  has studied the strong decays of heavy baryons in heavy hadron chiral perturbation theory(HHChPT). The decay processes $\Sigma _b^*  \to \Lambda _b \pi$ and $\Sigma
_b \to \Lambda _b \pi$ have been studied by combining the chiral
dynamics and the MIT bag model \cite{Hwang}.

The decay width is computed as \cite{Fayyazuddin1997}
\begin{eqnarray}
\Gamma(\Sigma_{b}\rightarrow
\Lambda_{b}\pi)= \frac{g^{2}_{A}}{2 \pi F^{2}_{\pi}} p^{3}_{\pi} \nonumber\\
\Gamma(\Sigma^{*}_{b}\rightarrow
\Lambda_{b}\pi)= \frac{g^{*2}_{A}}{24 \pi F^{2}_{\pi}} p^{3}_{\pi} \frac{(m_{\Sigma^{*}_{b}}+ m_{\Lambda_{b}})^{2}}{m^2_{\Sigma^{*}_{b}}}
 \end{eqnarray}
Where, $g_{A}$ and $g^{*}_{A}$ are the axial vector coupling constants for the transitions $\Sigma_{b}\rightarrow \Lambda_{b}\pi$
and $\Sigma^{*}_{b}\rightarrow \Lambda_{b}\pi$ respectively and is taken as $g_{A}=-\sqrt{\frac{2}{3}}g^{N}$
 and $g^{*}_{A}=-\sqrt{2} g_{A}$ \cite{Fayyazuddin1997}. Using the experimental value $g^{N}_{A}=1.25$, $F_{\pi}=130$ MeV and $p_{\pi}$,the momentum carried by the pion computed as
$ \sqrt{(\frac{m^{2}_{\Sigma_{b}}-m^{2}_{\Lambda_{b}}}{2 m_{\Sigma_{b}}})^{2}-m^{2}_{\pi}}$, the decay widths are obtained. The computed decay widths are listed in Table \ref{tab:06} along with other model predictions.

\begin{table}[t] \caption{\label{tab:06} Strong decay width of the $\Sigma^{*}_{b}$ and
$\Sigma_{b}$ Baryons in terms of MeV.}
\begin{center}
\begin{tabular}{lrrrrrr}

\hline
Parameters &$\nu$&$\Gamma_{(\Sigma_{b}\rightarrow \Lambda_{b}\pi)}$&$\Gamma_{(\Sigma^{*}_{b}\rightarrow
\Lambda_{b}\pi)}$\\
\hline

            &	0.4	& 1.07&5.02\\
        	&	0.6	&  1.71&5.89\\
	 Set A	&	0.8	&2.20&6.50\\
	      	&	1.0	& 2.47&6.82\\
\hline
            &	0.2	&1.95&5.87\\
            &	0.4&3.33&7.40\\
        	&	0.6	& 4.63&8.74\\
    Set B   &	0.8&4.63&8.74\\
	      	&	1.0	&4.63&8.74\\
\hline

Others \cite{Xin2008} &&6.73-13.45   & 10.00-17.74 \\

         \cite{Hwang} &&4.35-5.77   &  8.50-10.44   \\
         \cite{A. Limphirat}  &&8.00&     15.00  \\
         \hline

\end{tabular}
\end{center}
\end{table}
\section{Semileptonic decay of $\Lambda_{b}\rightarrow \Lambda_{c}l\nu_{l}$}
Among all the processes, the semileptonic decay of hadrons plays
an important role for probing the success and predictability of the model
that describes the hadron state. This decay process is relatively simple and less
dependent on the non-perturbative QCD effects as the leptons do
not participate in strong interaction. And there is no
contamination from the crossed gluon-exchanges between quarks
residing in different hadrons which are produced in the weak
transitions. Thus one might gain more model-independent
information, such as extraction of the Cabibbo-Kobayashi-Maskawa
matrix elements from semileptonic decay rates. In the semi-leptonic decays of heavy
hadrons, it is expected to factorize the perturbative and
non-perturbative parts more naturally. Recently, semileptonic decay of $\Lambda_{b}$ has been reported in the light-front quark model and diquark picture \cite{Hong2008}.\\
The inclusive semileptonic decay width of the beauty baryon $\Lambda_{b}\rightarrow \Lambda_{c}l\nu_{l}$ is
given by \cite{Korner2000,Nir1989}

\begin{center}
\begin{equation}
\Gamma^{inc}= \Gamma_{0} (1-\frac{2}{3}\frac{\alpha_{s}}{\pi} g(x))
\end{equation}
\end{center}
Where, $x = (m_c/m_b)^2$ and $\Gamma_{0}$ is the lowest order free quark decay and is expressed as \cite{Korner2000}
\begin{center}
\begin{equation}
\Gamma_{0}= \frac{G^{2}_{F} |V_{cb}|^{2} m^{5}_{b}} {192  \pi^{3}} I_{0}(x)
\end{equation}
\end{center}
the function  $g(x)$ is written as \cite{Korner2000,Nir1989}
\begin{equation}
g(x)=h(x)/I_0(x)\; , \nonumber  \\
\end{equation}
Where,
\begin{eqnarray}
h(x)= -(1-x^2)\left ( \frac{25}{4} - \frac{239}{3} x + \frac{25}{4} x^2 \right )
+ x \ln(x)\left (20 + 90 x - \frac{4}{3} x^2 + \frac{17}{3} x^3 \right )\nonumber  \\ +
x^2 \ln^2(x)
(36 + x^2) +  (1-x^2)\left (\frac{17}{3} - \frac{64}{3} + \frac{17}{3}x^2 \right ) \ln
(1-x) \nonumber  \\- 4 (1 + 30 x^2 + x^4)\ln (x) \ln (1-x) \nonumber \\
- (1 + 16 x^2 + x^4) [ 6 {\rm Li}_2 (x) - \pi^2 ] \nonumber \\
- 32 x^{3/2} (1+x) \left [ \pi^2 - 4 {\rm Li}_2 (\sqrt{x}) +
4 {\rm Li}_2( -\sqrt{x}) - 2 \ln (x) \ln \left (
\frac{1-\sqrt{x}}{1+\sqrt{x}} \right ) \right ]\nonumber \\
 and I_0(x) =  (1-x^2)(1-8 x + x^2) -12 x^2 \ln x \nonumber \\
\end{eqnarray}
Here, $Li_{2}(x)=\frac{x}{1}+\frac{x^{2}}{2^{2}}+\frac{x^{3}}{3^{2}}$ \\
As the decay width here depends on the mass (effective mass) of the heavy quarks within the intial ($\Lambda_{b}$)
 and final ($\Lambda_{c}$) baryons through the parameter, $x$, we compute the inclusive decay width with and without considering the bound state effect on the quark mass parameters. The semileptonic decay widths thus
 computed here are listed in Table \ref{tab:07} and are compared with other model predictions.
\begin{table*}
\begin{center}
\caption{\label{tab:07}Semileptonic decay widths ($\Gamma_{SL}$) of beauty baryons (* in $10^{10} s^{-1}$.)}
\begin{tabular}{lrrrrrrr}
\hline Decay& $\nu$&{\,\,\,\,\,\,\,\,\,\,\,\,\,\,\,\,\,\,\,\,}$\Gamma_{SL}$&& Others&\\
& &cqm&ecqm& &\\
\hline $\Lambda_{b}\rightarrow \Lambda_{c}l\nu_{l}$
&0.4&2.50&6.09&$3.59^{+1.234}_{-0.936}$\cite{K. Nakamura2010}-PDG\\
&0.6&2.97&6.22&5.90\cite{singl1991}\\
Set A&0.8&3.24&6.28&5.10\cite{ct1996}\\
&1.0&3.40&6.31&5.14\cite{kkp1994}\\
\cline{2-4}
&0.2&3.08&6.24&5.40\cite{hjkl2005}\\
&0.4&3.93&6.42\\
Set B&0.6&4.32&6.49\\
&0.8&4.55&6.54\\
&1.0&4.73&6.57\\
\hline
\end{tabular}
\end{center}
\end{table*}
\section{Results and Discussions}

The decay properties of single-beauty baryon have been studied based on the spectroscopic parameters of the baryon ground state. For the spectroscopic parameters of the baryon, we have employed a nonrelativistic quark-diquark model with coulomb plus power law interquark potential. The model parameters are
obtained to get the ground state spin average masses of the octet- decuplet (bqq) systems. The mass spiltting has been studied for different choices of the quark mass parameter, $m_{b}$ for each case of the potential exponent ($\nu$) and with different choices of $\alpha_{s}$. It is found that the value of heavy quark mass parameter and choice of $\alpha_{s}$ play a decisive role in the mass splitting of the ground state baryons. The resulting parameters for the different choices of the potential exponent $\nu$ are listed in Table \ref{tab:1} along with the predicted mass of $\Lambda_{b}$. Using these deduced spectroscopic mass parameters (Set A and Set B), we have computed the magnetic moments and decay properties of the octet and decuplet beauty baryons.\\
 The calculations of magnetic moments, the radiative transition widths and the semi-leptonic decay widths do depend explicitly on the quark mass parameter. Thus as an effective degrees of freedom for the confined quarks within the baryonic state, one must consider the confinement effect on the mass parameter of the quarks. We have computed these properties with constituent quark mass(cqm)parameter as well as by considering an effective mass of bound quarks inside the baryons(ecqm).
In general, our results on the magnetic moments with ecqm are in good agreement with other model predictions \cite{Amand2006}. Our results show lesser variations with change in the potential exponent,$\nu$.\\
Using the two sets of spectroscopic parameters (set A and B) and for the different choices of the potential exponenet, $\nu$, the transition magnetic moment and the decay width corrospond to $(\Sigma^{*0}_{b},\Sigma^{0}_{b})\rightarrow\Lambda_{b}\gamma$ and $\Sigma^{*}_{b}\rightarrow \Sigma_{b}\gamma$ transitions are computed and the results are listed in Table \ref{tab:04} and \ref{tab:05} respectively. Our results for the transition, $\Sigma^{*0}_{b}\rightarrow\Lambda_{b}\gamma$ lie in the range of 78-109 keV in the case of set A, while set B, the predictions lie in the range of 88-124 keV.\\ How ever, it is found that the result obtained with set B and with ecqm at $\nu=1.0$, is in agreement with the QCD sum rule prediction \cite{Aliev2008}. In the case of $\Sigma^{0}_{b}\rightarrow\Lambda_{b}\gamma$, our predictions lie in the range of 52-85 keV by considering both the sets together. No other model predictions are available for comparison. \\
In the case of $\Sigma^{*}_{b}\rightarrow \Sigma_{b}\gamma$, our predictions using either sets of the parameters are lower than other available model predictions. How ever, we find $\Gamma_{\Sigma^{*+}_{b}\rightarrow \Sigma^{+}_{b}\gamma} > \Gamma_{\Sigma^{*-}_{b}\rightarrow \Sigma^{-}_{b}\gamma} > \Gamma_{\Sigma^{*0}_{b}\rightarrow \Sigma^{+0}_{b}\gamma}$ in accordance with other theoretical predictions \cite{Zhu1999, Stawfiq2001, Aliev2008}.\\
The strong decay widths computed with the same set of spectroscopic parameters are listed in Table \ref{tab:06}.
The present results (6.19-9.22 MeV)for $(\Gamma_{\Sigma^{*}_{b}\rightarrow \Lambda_{b}\pi})$ obtained using the parameter Set B, are in excellant agreement with the MIT bag model predictions(8.50-10.44 MeV)\cite{Hwang}. \\
In the case of $\Sigma_{b}\rightarrow \Lambda_{b}\pi$, our predictions with set B lie in the range of (3.33-4.63) MeV for the choices of $\nu$ from 0.4 to 1.0. The predictions are found to saturate beyond $\nu \geq 0.6$. The saturated value of 4.63 MeV obtained here is well within the range of values, (4.35-5.77) MeV, predicted by the MIT bag model \cite{Hwang}. Other model predictions are found to be higher than the present values.\\
The semileptonic decay computed here with set of parameters (A and B) are listed with the experimental as well as with other available theoretical predictions. Our results with the parametric set B, without considering the effective quark mass (cqm) of the semileptonic decay width for $\Lambda_{b}\rightarrow \Lambda_{c}l\nu_{l}$ (3.08-4.73) $ *10^{10} s^{-1}$, are in good agreement with the experimental results \cite{K. Nakamura2010}, while those with the consideration of effective mass of the quark (ecqm) are closer to other theoretical predictions.\\

Though there are indications on the importance of the confinement effect on the quark mass parameters in the successful predictions of the various properties (magnetic moments, electromagnetic decay width, semileptonic decay width) of the ocetet - decuplet single beauty baryon, there is lack of sufficient experimental data corresponding to some of these properties studied here. However, our results of the semileptonic decay of $\Lambda_{b}\rightarrow \Lambda_{c}l\nu_{l}$ without considering the effective mass of the confined quarks are well within the experimental error bar. We look forward to more experimental data related to the beauty baryons coming from the ongoing as well as the upcoming facilities.
\\\\

\section{Acknowledgements} We acknowledge the financial support from University Grant Commission,
Government of India, under a Major Research Project \textbf{F. 32-31/2006(SR)}.\\

\end{document}